# Magnetism of Fullerene $C_{60}$ Compared with Graphene Molecule by DFT Calculation, Laboratory Experiment and Astronomical Observation

Norio Ota[1], Aigen Li[2], Laszlo Nemes[3] and Masaaki Otsuka[4]

[1]Graduate school of Pure and Applied Sciences, University of Tsukuba, *1-1-1 Tennodai, Tsukuba-City Ibaraki, 305-8571, Japan*
[2]Department of Physics and Astronomy, University of Missouri, *Columbia, MO 65211, USA*,
[3]Research Center for Natural Sciences, Ötvös Lóránd Research Network, *Budapest 1519, Hungary*
[4]Okayama Observatory, Kyoto University, *Asakuchi Okayama, 719-0232, Japan*.

Magnetism of fullerene $C_{60}$ was studied by three methods of the density functional theory (DFT) calculation, laboratory experiment and astronomical observation. DFT revealed that the most stable spin state was non-magnetic one of $Sz=0/2$. This is contrary to our recent study on void induced graphene molecules of $C_{23}$ and $C_{53}$ to be magnetic one of $Sz=2/2$. Two graphene molecules combined model suggested that two up-spin at every carbon pentagon ring may cancel each other to bring $Sz=0/2$. Similar cancelation may occur on $C_{60}$. Molecular vibrational infrared spectrum of $C_{60}$ show four major bands, which coincide with gas-phase laboratory experiment, also with astronomically observed one of carbon rich planetary nebula Tc1 and Lin49. However, there remain many unidentified bands on astronomical one. We supposed multiple voids on graphene sheet, which may create both $C_{60}$ and complex graphene molecules. It was revealed that spectrum of two voids induced graphene molecule coincident well with major astronomical bands. Simple sum of $C_{60}$ and graphene molecules could successfully reproduce astronomical bands in detail.

**Key words**: fullerene C60, graphene, DFT, gas-phase experiment, planetary nebula, infrared spectrum

## 1. Introduction

Graphene and graphite like carbon materials are candidates for showing ferromagnetic like hysteresis[1)-6)]. There are many capable explanations based on impurities[7)], edge irregularities[8)-10)] or defects[11)-17)]. The density functional theory (DFT) shows good coincidence with experiments[18)-21)] and revealed reasonable consistency with several fundamental theories[22)-23)]. Despite such many efforts, origin of magnetic ordering could not be thoroughly understood. Our recent studies on graphene nano-ribbon (GNR)[17)] and on graphene molecules[24)] revealed that void-defect brings unusual highly polarized spin configuration and structure change. Especially on graphene molecules[24)], we could firstly suggest good coincidence of DFT calculated molecular vibrational infrared spectrum both with astronomical observation of carbon rich planetary nebula[25)], and with laboratory experiment of laser induced carbon plasma[26)]. Model molecules for DFT calculation were graphene molecules of $C_{23}$ and $C_{53}$ induced by single void-defect. They have one carbon pentagon ring combined with many hexagon rings. Single void has three radical carbons, holding six spins. Those spins make several spin-states, which affects to molecular structure and molecular vibration, finally to infrared spectrum.

Corresponding author: Norio Ota (n-otajitaku@nifty.com).

It was amazing that the stable spin state was magnetic one of spin state $Sz=2/2$, not non-magnetic one of $Sz=0/2$. Here, there is an interesting question that fullerene $C_{60}$ is magnetic or not. Fullerene has 12 carbon pentagon rings combined with 20 hexagons, which is similar combination with void-induced graphene molecules. In this paper, we like to study magnetism of $C_{60}$ by three methods of DFT calculation, laboratory experiment[27), 28)]. and astronomical observation[25)]. Comparison with astronomical infrared spectrum is important that DFT calculates isolate and zero temperature status molecule. Also, astronomical carbon dust is similar situation, floating in interstellar space under ultra-low density (1~100 molecules/cm$^3$) and low temperature (few to 10 K) conditions.

Fullerene $C_{60}$ was discovered by Kroto et al.[30)] (the 1996 Nobel Prize) in the sooty residues of vaporized carbon. They already suggested that "fullerene may be widely distributed in the universe". The presence of $C_{60}$ in astrophysical environments was revealed by the detection of a set of emission bands at 7.0, 8.45, 17.3 and 18.9 µm[31)-37)]. Typical astronomical objects are the Galactic planetary nebula (PNe) Tc1[31)] and the Small Magellanic Cloud PNe Lin49[25)]. Observed spectra were compared with experiment and theory[38)-40)]. However, there remain undetermined observed bands, cannot be explained only by $C_{60}$.

It is well known that graphene is a raw material for synthesizing fullerene[39), 40)]. By observation of Lin49, Otsuka et al.[25)] suggested the presence of small graphene. Graphene was first experimentally synthesized by Geim and Novoselov[41] (the 2010 Nobel Prize). The possible presence of graphene in space was reported by Garcia-Hernandez et al.[42-44]. These infrared features appear to be coincident with planar $C_{24}$ having seven carbon hexagon rings. However, full observed bands still cannot be explained by $C_{24}$ or hexagon network molecules[45]. Some hints come from carbon SP3 defect among SP2 network caused by void-defect in carbon hexagon network by Ota[46]. Also, Galue & Leines[47] predicted the physical model supposing pi-electron irregularity. Resulted molecular configuration was carbon pentagon rings among hexagon networks.

At first part of this paper, stable spin state and vibrational infrared spectrum of $C_{60}$ will be calculated and compared with laboratory experiments and astronomical observation. Those will be compared with single-void induced graphene $C_{23}$ and $C_{53}$. At second part, multiple-void induced graphene molecules as like $C_{22}$, $C_{21}$, $C_{52}$ and $C_{51}$ will be calculated, which results are compared with that of fullerene $C_{60}$.

## 2. Calculation Method

In calculation, we used DFT[48)49)] with the unrestricted B3LYP functional[50]. We utilized the Gaussian09 software package[51] employing an atomic orbital 6-31G basis set[52]. Unrestricted DFT calculation was done to have the spin dependent atomic structure. The required convergence of the root-mean-square density matrix was $10^{-8}$. Based on such optimized molecular configuration, fundamental vibrational modes were calculated, such as carbon to carbon (C-C) stretching modes, C-C bending modes and so on, using the Gaussian09 software package. This calculation also gives harmonic vibrational frequency and intensity in infrared region. The standard scaling is applied to the frequencies by employing a scale factor of 0.975 for pure carbon system taken from the laboratory experimental value of 0.965 based on coronene-molecule of $C_{24}H_{12}$ [46]. Correction due to anharmonicity was not applied to avoid uncertain fitting parameters. To each spectral line, we assigned a Gaussian profile with a full width at half maximum (FWHM) of 4cm$^{-1}$.

It should be noted that any molecular symmetricity was not applied to compare delicate change of molecular configuration and compare small difference of total energy.

## 3. Spin State Analysis of Fullerene $C_{60}$

### 3.1 Stable spin state of $C_{60}$ compared with graphene molecule.

For DFT calculation of $C_{60}$, input condition is charge neutral for two cases of spin state of $Sz$=0/2 or $Sz$=2/2. In this study, we dealt total molecular spin $S$ (vector). Molecule is rotatable material, easily follows to the external magnetic field of z-direction. Projected component $Sz$ to z-direction is good quantum number.

Result is shown in Fig. 1. Total molecular energy of $Sz$=0/2 is 1.79 eV lower (stable) than that of $Sz$=2/2, that is, $C_{60}$ is not magnetic and show no spin-distribution. Spin-density of $Sz$=2/2 was mapped on right bottom, of which up-spin cloud (by red) surrounds around an equator of soccer ball like $C_{60}$. Such excess up-spin cloud of $Sz$=2/2 brings slight energy increase than $Sz$=0/2.

Stable spin state of $C_{60}$ was contrary to our previous study on single-void-defect induced graphene molecules of $C_{23}$ which show stable spin state to be $Sz$=2/2 [24]. Typical example was illustrated in Fig. 2. Starting molecule was $C_{24}$ having seven carbon hexagon rings. High-speed particle attacks one carbon atom and kick out. Single void will be created. DFT calculation gives optimized molecular configuration. Resulted molecule was $C_{23}$ having one pentagon ring. In our previous paper[24], we named ($C_{23}$-c) to distinguish void position. Initial void-defect holds 3 radical carbons and allows 6 spins. Six spins make capable spin-states of $S_z$=0/2, 2/2, 4/2 and 6/2. Among them, calculated energy of $S_z$=4/2 and 6/2 resulted unstable high energy. Here, we should compare cases of $S_z$=0/2 and 2/2. It should be noted that molecular energy of $Sz$=2/2 was 0.64 eV lower (stable) than that of $Sz$=0/2. We can see up-spin major spin cloud for $Sz$=2/2 on bottom right at a cutting surface of spin density at 10e/nm$^3$.

Molecular vibrational infrared bands of $C_{60}$ were shown in Table 1, for $Sz$=0/2 in column (d) and for $Sz$=2/2 in (e). Fundamental vibrational mode of $Sz$=0/2 were noted in (d). We noticed four major bands. Mode of Band-A (18.72 µm) was C-C stretching on $C_{60}$ surface fixing carbon to carbon length of pentagon ring. Mode of Band-B (17.44 µm) was asymmetric spherical motion of $C_{60}$ ball. Modes of Band-C (8.58 µm) and Band-D (6.77 µm) were both C-C stretching on $C_{60}$ surface.

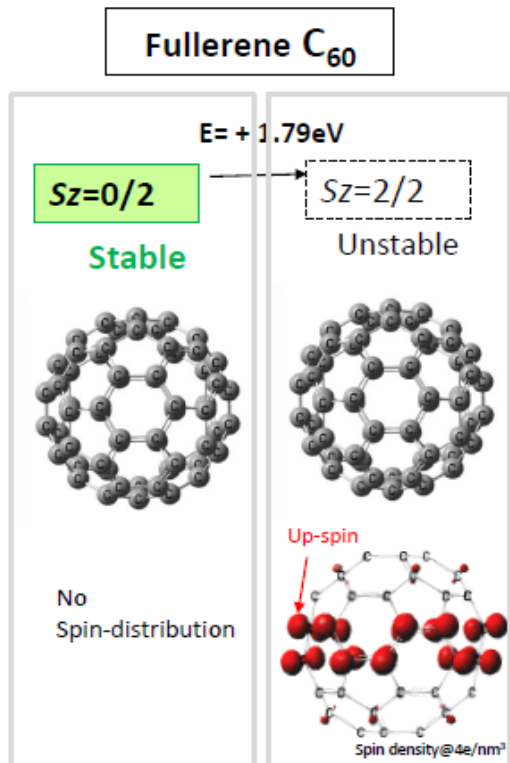

**Fig. 1** Spin state of fullerene $C_{60}$ by DFT calculation.

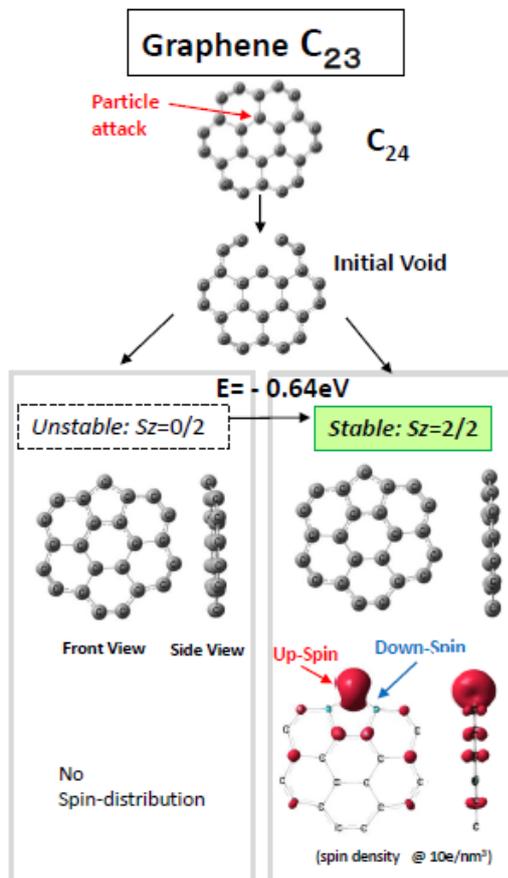

**Fig. 2** Spin state of single void induced graphene $C_{23}$.

### 3.2 Spin alignment of $C_{60}$

It should be noted that stable spin state of $C_{60}$ is $Sz=0/2$. This is contrary to that of $C_{23}$ and $C_{53}$. To understand such difference, we consider simple spin alignment model as shown in Fig. 3. Two molecules of $C_{23}$ were supposed as illustrated on column (A). As shown on right, if two carbon atoms (1C, 2C) of each carbon pentagon rings (marked by circled 5) were far enough, there remains original up-spin cloud as up-up pair. Coupled molecules may show total spin of $Sz=4/2$. Whereas as shown on left, two molecules were close enough, two molecules will make a new bond. Spin alignment will be up-down pair to result $Sz=0/2$. Such new covalent bond will reduce total energy. Under such consideration, we can suggest two capabilities of $C_{60}$ as shown on panel (B), that is, non-magnetic one on left and magnetic one on right. By DFT calculation, we could obtain precise molecular configuration on (C) and could judge preferable spin state to be $Sz=0/2$.

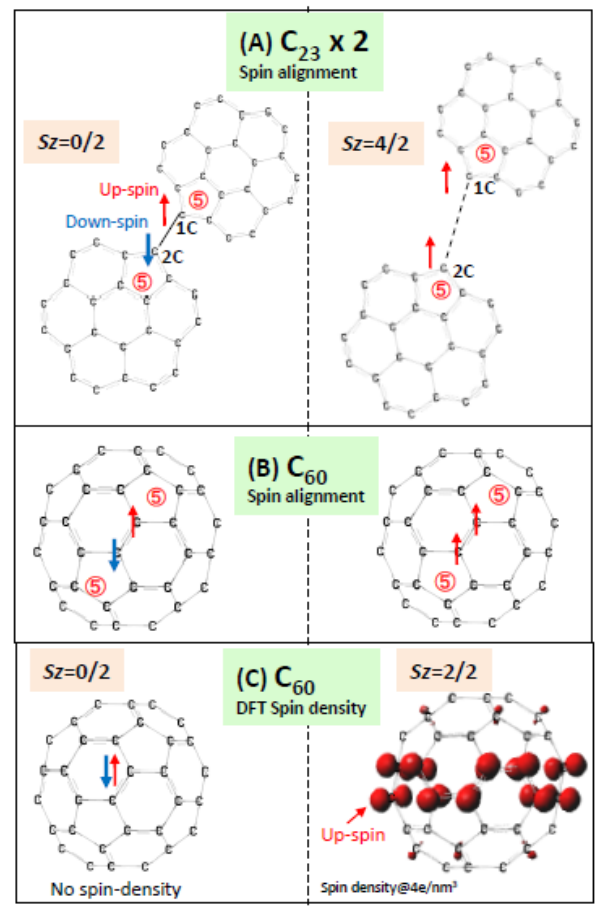

**Fig. 3** Spin alignment model by two graphene molecules of $C_{23}$ was illustrated in (A), analogy to $C_{60}$ in (B). DFT calculated spin distribution of $C_{60}$ was shown in (C).

**Table 1** Molecular vibrational bands of fullerene $C_{60}$.

| | C60: Laboratory Experiment | | | C60: DFT Calculation | |
|---|---|---|---|---|---|
| | (a) Gas-phase (~1000 K) [Nemes et al.] | (b) Gas-phase (Extrapolate to 0 K) [Nemes et al.] | (c) Solid-phase (300 K) [Kratschmer et al.] | (d) Sz=0/2 (Stable) [This study] | (e) Sz=2/2 (Unstable) [This study] |
| Remarks | Gas-phase experiment | Extrapolation to 0 K | Solid-phase experiment | Isolate C60 at zero-temperature | |
| Material phase | (Gas-phase, 1000-> 0 K) | | (Solid-phase, 300 K) Silicon substrate | (Isolate, 0 K) | |
| Band-A | 18.95 | 18.82 | 18.93 | 18.72 (C-C stretching on C60-surface fixing pentagon ring's carbon to carbon length) | 18.71 |
| Band-B | 17.54 | 17.41 | 17.33 | 17.44 (Asymmetric spherical motion of C60 ball) | 17.68 |
| Infrared band (micrometer) | | | | | 13.71 |
| | | | | | 13.35 |
| Band-I | 8.58 | 8.40 | 8.45 | 8.58 (C-C stretching on C60-surface) | |
| Band-L | 7.11 | 6.97 | 7.00 | 6.77 (C-C stretching on C60-surface) | 7.00 |

**Table 2** Vibrational bands of graphene molecules.

| Laboratory Experiment | | | Astronomical Observation | | DFT Calculation (0 K, Isolate molecule) | | | | | | | |
|---|---|---|---|---|---|---|---|---|---|---|---|---|
| C60 Gas-phase (Extrapolated to 0 K) | C60 Solid (300K) | Laser induced carbon (300K) | Tc1 Nebula | Lin49 Nebula | C60 Sz=0/2 | C24 Sz=0/2 | C23 Sz=2/2 | C22 Sz=2/2 | C21 Sz=4/2 | C53 Sz=2/2 | C52 Sz=2/2 | C51 Sz=2/2 |
| Number of carbon rings hexagon: 20 pentagon: 12 | − − | | | | 20 12 | 7 0 | 6 1 | 5 2 | 4 3 | 18 1 | 17 2 | 16 3 |
| (Wavelength in micrometer) | | | | | 20.1 | | | 19.8 | | | 19.6 | 19.1 |
| | | | | | | | 19.1 | 19.2 | | | | |
| | 18.93 | | 18.9 μm | 18.9 | 18.72 | | 18.9 | 19.0 | | 18.9 | 18.9 | |
| 18.82 | | | | | | | | | | | | 18.6 |
| 17.41 | 17.33 | | 17.4 | 17.4 | 17.44 | | 17.4 | 17.3 | 17.4 | | | |
| | | | | 16.6 | | | 16.9 16.5 | 16.4 | | | | |
| | | | | 14.3 | | | 14.3 | | | 14.3 | | |
| | | | | 13.9 | | | | | | 13.9 | 13.8 13.5 | |
| | | | | 13.3 | | | | | 12.7 | 13.4 | 13.3 11.4 | 13.1 |
| | | 10.2 9.5 | | 10.0 | | 9.8 | 9.9 | 10.2 | | | 9.3 | |
| | | | 9.0 | | | | 9.0 | 8.9 | 8.9 | 8.6 | | 9.0 |
| 8.40 | 8.45 | 8.3 | 8.4 | 8.4 | 8.58 | | | | | | 8.4 | 8.3 |
| | | | | 8.1 | | | 8.0 | 8.1 7.9 | 7.9 | | | |
| | | 7.7 | 7.6 | 7.6 | | | 7.5 | 7.6 | | 7.6 | 7.6 | 7.6 |
| | | 7.4 | | | | | 7.4 | | | 7.3 | | |
| 6.97 | 7.00 | Plateau (5.9–7.7) | 7.1 | 7.1 | | | 7.1 | 6.9 | 7.2 | 7.1 | 7.1 | 7.0 |
| | | 6.7 | 6.5 | 6.6 | 6.77 | 6.6 | 6.6 | | | 6.5 6.3 | 6.3 | 6.5 |
| | | 5.9 | | | | | | | | | | |

### 3.3 Comparison with laboratory experiments

There are several laboratory experiments of infrared emission bands of $C_{60}$ as summarized on Table 1. In a column (c), emission bands of solid phase $C_{60}$ clusters, sticked on silicon substrate, was measured at 300 K by Kratschmer et al.[28], which show four bands at 18.93, 17.33, 8.45 and 7.00 µm. Solid-phase experiment has an advantage to give sufficient strong emission signal. DFT calculated bands (d) were close to those bands within 0.2 µm discrepancy. However, we should mind that molecular motion of solid phase would be bound by neighbor molecules and by silicon substrate. Gas-phase experiments is necessary to obtain near isolate and motion free emission spectrum. Nemes et al.[27] show high temperature (879~1212 K) emission bands as noted in (a) to be 18.95, 17.54, 8.58, and 7.11µm, which show longer wavelength than solid-phase one. To compare DFT calculation, low temperature behavior is necessary, but emission intensity is so weak. They simply extrapolated to 0 K based on high temperature experiment as shown in (b), where Band-A, is 18.82 µm close to DFT obtained 18.72 µm (d). Band-B just coincide well each other at 17.4µm. Band-C show 8.40 µm (b) compared to DFT calculated 8.58 µm (d). Also, Band-D was 6.97 µm (b) compared to calculated 6.77 µm (d). They show reasonable coincidence within 0.2 µm discrepancy.

### 3.4 Comparison with astronomically observed infrared-spectrum

Comparison of DFT calculation with astronomically observed bands is important, because they are both ideal cases of ultra-low material density and at low temperature. Otsuka et al.[25] opened astronomically observed bands as illustrated in Fig. 4, on top for Tc1 nebula by red and for Li49 nebula by blue. On middle, gas-phase zero-temperature extrapolated bands of $C_{60}$ were marked by blue arrows. DFT calculated spectra were illustrated on bottom for the case of $Sz$=0/2 by green and for $Sz$=2/2 by light green. It should be noted that the observed spectra are seen in emission. A nearby star as like a central star of Tc1 nebula may illuminate the molecules and excites them to give rise infrared emission. Detailed discussion was done by Li and Drain[53],[54]. We regard that DFT calculated absorbed spectrum is mirror image of emission one in case of sufficient large photon energy excitation due to the theory of Einstein's emission coefficient[53],[54].

Calculated spectrum of $Sz$=2/2 (by light green) was so different with astronomical one. On the other hand, spectrum of $Sz$=0/2 (by green) looks good to reproduce astronomical one. Band behavior is ambiguous for $Sz$=2/2. Energy favorable spin state of $Sz$=0/2 show specific four bands. Both Tc1 and Lin49 show major band at 18.9 µm, which is only 0.1 µm longer than gas-phase experimental one, and 0.2 µm longer than DFT calculated one of $C_{60}$. It looks fair coincidence. Sharp emission at 18.7 µm (marked by black arrow) is a strong atomic emission line of $S_{III}$. We are afraid that such strong atomic emission may hide true molecular band of $C_{60}$. DFT calculated Band-B (17.44 µm) coincides so well both with astronomically observed one and with gas-phase laboratory experimental one. In shorter wavelength region, calculated $C_{60}$ has two bands at 8.58 and 6.77 µm. It looks fair coincidence with gas-phase experimental bands at 8.40 and 6.97 µm, also with astronomically observed one at 8.4 and 6.6 µm.

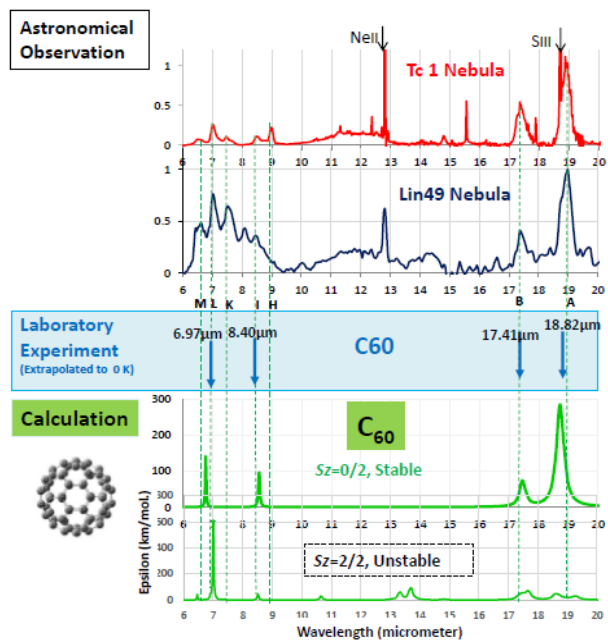

**Fig. 4** Comparison of molecular vibrational infrared spectra on fullerene $C_{60}$ by astronomical observation, by laboratory experiment and by DFT spin dependent calculations.

It should be noted that there are many unidentified astronomical bands. Those complex bands may come from unknown carbon molecules, not only by single molecule of $C_{60}$. Our previous study[24] considered single void-defect on mother graphene of $C_{24}$ and $C_{54}$. Resulted graphene were $C_{23}$ and $C_{53}$, which infrared spectra are again illustrated in Fig. 5 and Table 2. Major band of $C_{23}$ has twin peaks at 18.9 and 19.0 µm, also $C_{53}$ has single peak at 18.9 µm, which are related to observed Band-A. Second major observed band was at 17.4 µm (Band-B), which was reproduced well by $C_{60}$. Unfortunately, Band-B was reproduced by $C_{23}$ and $C_{53}$ only by weak intensity band. At shorter wavelength region from 6 to 10 µm, spectrum of $C_{60}$ only show two bands of Band- I and –L, while $C_{23}$ and $C_{53}$ could reproduce many bands as like G, H, I, J, K, L and M. Moreover, $C_{53}$ could reproduce detailed bands of D, E, F. In Table 2, Nemes et al. reported another laboratory experiment by the laser induced carbon plasma[24),26)]. We can see rough coincidence with bands from graphene molecules, little coincidence with $C_{60}$.

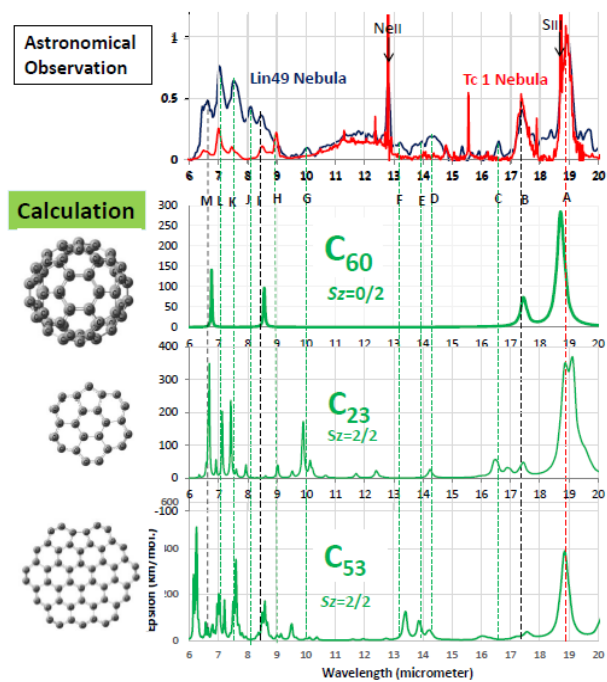

**Fig. 5** DFT calculated molecular vibrational spectra of ($C_{60}$, $Sz=0/2$), ($C_{23}$, $Sz=2/2$) and ($C_{53}$, $Sz=2/2$) compared with astronomically observed spectra of carbon rich nebulae Tc1 and Lin49.

## 4. Multiple-Void Induced Carbon Molecules

### 4.1 The top-down process

It should be noted that infrared spectrum of $C_{60}$ show strong band at 17.4 μm (Band-B), whereas single-void induced $C_{23}$ and $C_{53}$ do not as compared in Table 2 by a red frame. Fullerene $C_{60}$ includes 12 pentagon rings, which suggested us that Band-B may come from multiple pentagon rings. Most chemists think that $C_{60}$ will be synthesized from small carbon molecules to larger one under heat and catalyst chemical reactions as illustrated on bottom of Fig. 6 by blue panel, which is so called the bottom-up process. Here, we consider reversed process called the top-down process as illustrated in red panel. The top-down process should start from large graphene sheet. This will be attacked by many high-speed particles (red arrows) to create multi-voids. Finally, a graphene sheet transforms to complex graphene molecule having a lot of carbon pentagon rings among hexagon networks. Fullerene $C_{60}$ may be typical example of such multiple voids induced molecule. Top-down process could be done under low temperature and pure physical quantum-mechanical procedure, which may be popular in space.

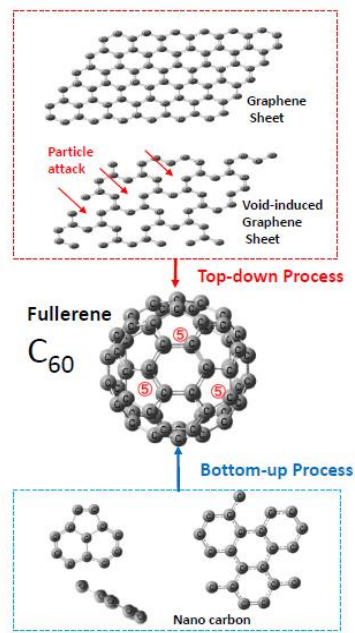

**Fig. 6** Top-down and bottom-up process of $C_{60}$.

### 4.2 Stable spin state of multi-void induced graphene

To understand the top-down process, we tried simple model starting from $C_{24}$. Results were summarized in Fig. 7. A single void induced graphene of $C_{23}$ has one carbon pentagon ring as marked by circled 5. Particle attacks a carbon at red arrowed site. Most stable spin-state was $Sz=2/2$. Two-voids induce $C_{22}$ having two carbon pentagon rings. Molecular configuration was umbrella like one (see side view). Possible spin states are $Sz=0/2$, $2/2$ and $4/2$. Most stable state was $Sz=2/2$, which energy was 0.17 eV lower than that of $Sz=4/2$ and 1.06 eV lower than $Sz=0/2$. Spin-distribution was illustrated on right. Top carbon of every pentagon ring was up-spin major. Similarly, three voids create $C_{21}$ having triple carbon pentagons. Stable spin state was $Sz=4/2$.

Size dependence was checked. Large size molecules starting from $C_{54}$ were analyzed as shown in Fig. 8. Single void induced $C_{53}$ has stable spin state of $Sz=2/2$. Two voids induced $C_{52}$ having $Sz=2/2$. Three voids induced $C_{51}$ having $Sz=2/2$.

In Fig.9, DFT calculated spectra for ($C_{23}$, $Sz=2/2$), ($C_{22}$, $Sz=2/2$) and ($C_{21}$, $Sz=4/2$) were compared. It was a surprise that Band-B (red broken line) was reproduced well by $C_{22}$ and $C_{21}$ as the largest band. Astronomically observed Band-B may come from those multi-pentagon rings.

Also in Fig. 10, calculated spectra of $C_{54}$ family are compared. Major band was just 18.9 μm for both $C_{53}$ and $C_{52}$. In case of $C_{51}$, the largest band was 19.2 μm. However, we could not find any large intensity at 17.4 μm of Band-B. Ratio of number of carbon pentagon rings to hexagons may contribute to the intensity

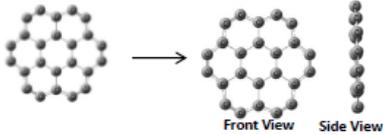

**Fig. 7** Stable spin state of graphene $C_{24}$ family.

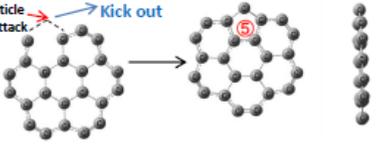

**Fig. 8** Stable spin state of $C_{54}$ family.

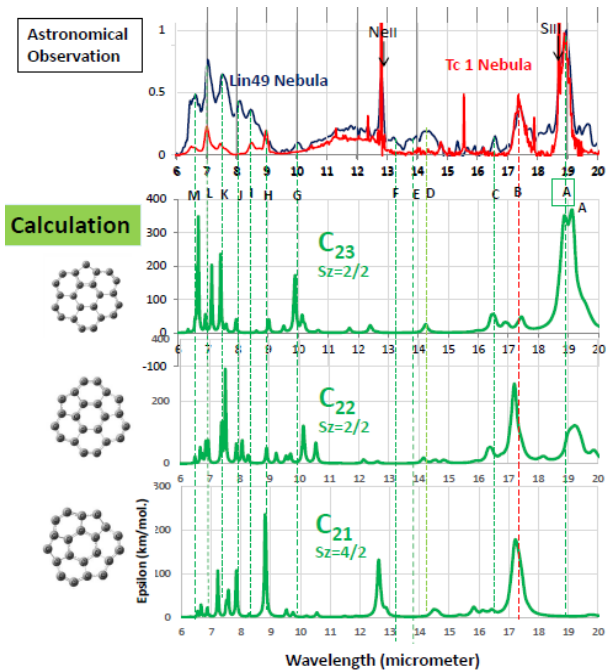

**Fig. 9** DFT calculated spectra of ($C_{23}$, $Sz=2/2$), ($C_{22}$, $Sz=2/2$) and ($C_{21}$, $Sz=4/2$) compared with astronomically observed ones.

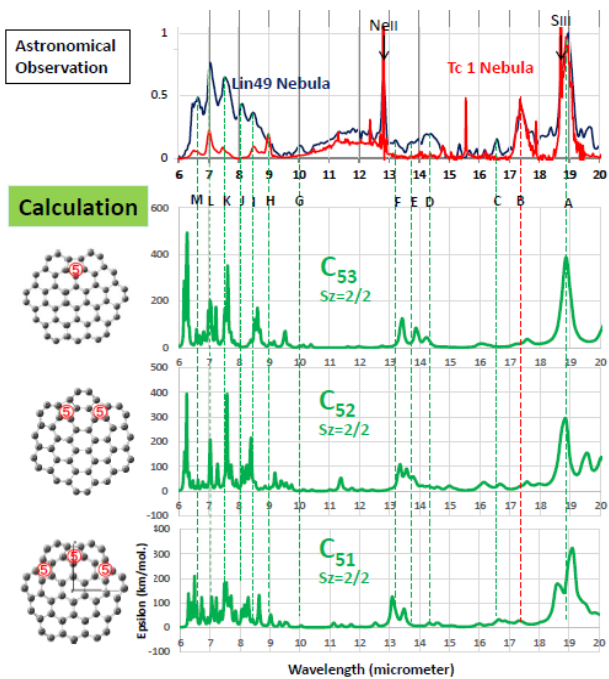

**Fig. 10** DFT calculated spectra of ($C_{53}$, $Sz=2/2$), ($C_{52}$, $Sz=2/2$) and ($C_{51}$, $Sz=2/2$) compared with astronomically observed ones.

## 5. Reproduction of astronomically observed spectrum.

We tried to reproduce astronomically observed spectrum using DFT spectra of $C_{60}$ and graphene molecules. As shown in Fig. 11, observed spectrum of Tc1 in panel (A) could be reproduced fairly well by $C_{60}$ in panel (C). It should be minded that Band-H and K of Tc1 could not reproduced by $C_{60}$. Observed spectrum of Lin49 in (B) is so complex, could not be reproduced only by $C_{60}$. Here, we obtained simple weighting sum of DFT calculated spectrum of graphene molecules of $C_{23}$, $C_{22}$, $C_{53}$, and $C_{52}$ as shown in (D). Observed major bands of Band-A and B were well reproduced. Also detailed bands of Band-C, D, E, F, were reproduced well. Further, shorter wavelength bands (Band-G, I, K, L and M) were also well reproduced again. Here, we add a DFT spectrum of $C_{60}$ to graphene as illustrated in (E). It looks that major part are similar with (D), may come from graphene molecules.

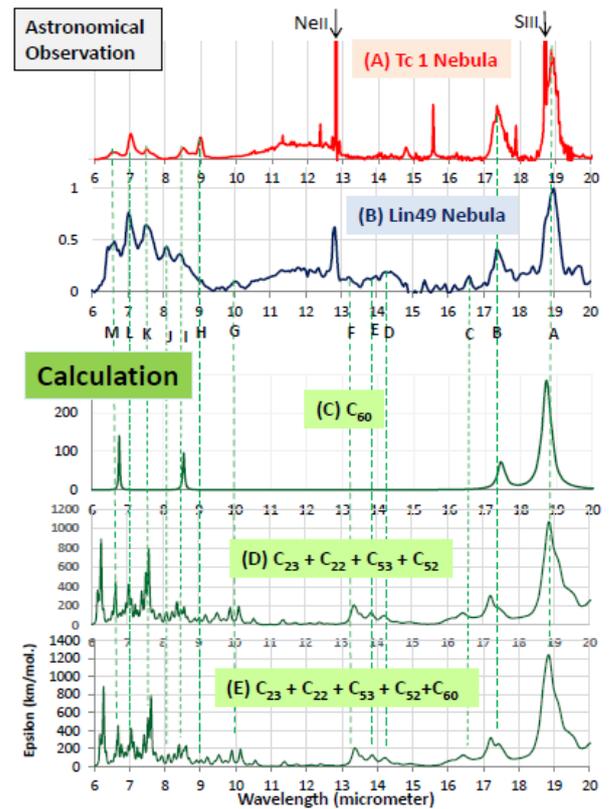

**Fig. 11** DFT calculated molecular vibrational spectrum is shown compared with astronomical observation of Tc1 nebula on (A) and Lin49 on (B). DFT calculated spectrum of $C_{60}$ was shown in (C). Simple sum spectrum of $C_{23}$, $C_{22}$, $C_{53}$ and $C_{52}$ is illustrated on (D). Further addition of $C_{60}$ is on (E).

## 5. Conclusion

Stable spin state and molecular vibrational infrared spectrum of fullerene $C_{60}$ was studied comparing with void induced graphene molecules.

(1) Spin state of fullerene $C_{60}$ were calculated by DFT to result that stable one was non-magnetic of $Sz=0/2$. This is contrary to graphene molecules of $C_{23}$ and $C_{53}$ to be magnetic $Sz=2/2$.

(2) Spin alignment model by two graphene molecules suggested that up-spin on each carbon pentagon ring cancel each other to bring up-down spin pair to be non-magnetic $Sz=0/2$. There may occur similar cancelation in $C_{60}$.

(3) Simulated molecular vibrational infrared spectrum of $C_{60}$ show four modes, which coincide well with laboratory experiment of gas-phase emission spectrum. Also, those coincide well with astronomically observed bands of nebula Tc1 and Lin49.

(4) There remain unidentified astronomical bands. Multiple voids induced graphene were studied. Two voids induced $C_{22}$ having two pentagon rings could reproduce remained and detailed astronomical observation.

(5) Simple sum of DFT calculated spectra by $C_{60}$ and graphene molecules ($C_{23}$, $C_{22}$, $C_{53}$ and $C_{52}$) could successfully reproduce astronomical bands in detail.

**Acknowledgement** Aigen Li is supported in part by NSF AST-1311804 and NASA NNX14AF68G.

## References

(Note on abbreviation of astronomical journals,
ApJ: The Astrophysical Journal
ApJL: The Astrophysical Journal Letters
A&A: Astronomy and Astrophysics
MNRAS: Monthly Notices of the Royal Astronomical Society
PNAS: Proceedings of the National Academy of Sciences)

1) P. Esquinazi, D. Spemann, R. Hohne, A.Setzer, K. Han, and T. Butz: *Phys. Rev. Lett.*, **91**, 227201 (2003).
2) K. Kamishima, T. Noda, F. Kadonome, K. Kakizaki and N. Hiratsuka: *J. Mag. Magn. Mat.*, **310**, e346 (2007).
3) T. Saito, D. Nishio-Hamane, S. Yoshii, and T. Nojima: *Appl. Phys. Lett.*, **98**, 052506 (2011).
4) Y. Wang, Y. Huang, Y. Song, X. Zhang, Y. Ma, J. Liang and Y. Chen: *Nano Letters*, **9**, 220 (2009).
5) J. Cervenka, M. Katsnelson and C. Flipse: *Nature Phys.*, **5**, 840 (2009), (https://doi.org/10.1038/nphys1399).
6) H. Ohldag, P. Esquinazi, E. Arenholz, D. Spemann, M. Rothermal, A. Setzer, and T. Butz: *New Journal of Physics*, **12**, 123012 (2010).
7) J. Coey, M. Venkatesan, C. Fitzgerald, A. Douvalis and I. Sanders: *Nature*, **420**, 156 (2002).
8) K. Kusakabe and M. Maruyama: *Phys. Rev. B*, **67**, 092406 (2003).
9) N. Ota, N. Gorjizadeh and Y. Kawazoe: *J. Magn. Soc. Jpn.*, **35**, 414 (2011), also **36**, 36 (2012).
10) N. Ota: *J. Magn. Soc. Jpn.*, **37**, 175 (2013).
11) P. Lehtinen, A. Foster, Y. Ma, A. Krasheninnikov, and R. Nieminen: *Phys. Rev. Lett.*, **93**, 187202 (2004).
12) P.Ruffieux, O. Groning, P. Schwaller, L. Schlapbach, and P. Groning: *Phys. Rev. Lett.*, **84**, 4910 (2000).
13) A. Hashimoto, K. Suenaga, T. Sugai, H.Shinohara, and S. Iijima: *Nature (London)*, **430**, 870 (2004).
14) K.Kelly and N.Hales: *Sur. Sci.*, **416**, L1085 (1998).
15) T. Kondo, Y. Honma, J. Oh, T. Machida, and J. Nakamura: *Phys. Rev. B*, **82**, 153414 (2010).
16) M. Ziatdinov, S. Fujii, K. Kusakabe, M. Kiguchi, T. Mori, and T. Enoki: *Phys. Rev. B*, **89**, 155405 (2014).
17) N. Ota and L. Nemes: *J. Magn. Soc. Jpn.*, **45**, 30 (2021).
18) K.Kelly and N.Hales: *Sur. Sci.*, **416**, L1085 (1998).
19) T. Kondo, Y. Honma, J. Oh, T. Machida, and J. Nakamura: *Phys. Rev. B*, **82**, 153414 (2010).
20) M. Ziatdinov, S. Fujii, K. Kusakabe, M. Kiguchi, T. Mori, and T. Enoki: *Phys. Rev. B*, **89**, 155405 (2014).
22) O. Yazyev and L. Helm: *Phys. Rev. B*, **75**, 125408 (2007).
23) B. Wang and S. Pantelides: *Phys. Rev. B*, **86**, 165438 (2012).
24) N. Ota, A. Li, L. Nemes and M. Otsuka: *J. Magn. Soc. Jpn.*, **45**, 41 (2021).
25) M. Otsuka, F. Kemper, M. L. Leal-Ferreira, M. L. Aleman, M. L. Bernard-Salas, J. Cami, B. Ochsendorf, E. Peeters, and P. Scicluna: *MNRAS*, **462**, 12 (2016).
26) L. Nemes, E. Brown, S. C. Yang, U. Hommerrich: *Spectrochimica Acta Part A, Molecular and Biomolecular Spectroscopy*, **170**, 145 (2017).
27) L. Nemes, R. S. Ram, P. F. Bernath, F. A. Tinker, M. C. Zumwalt, L. D. Lamb, D. R. Huffman: Chem. Phys. Lett., **218**, 295 (1994).
28) W. Kratschmer, L. D. Lamb, K. Fostiropoulos and D. R. Huffman: Nature, **347**, 354 (1990).
29) L. Nemes, A. Keszler, J. Hornkohl, and C. Parigger: *Applied Optics*, **44-18**, 3661 (2005)
30) H. W. Kroto, J. R. Heath, S. C. Obrien, R. F. Curl, and R. E. Smalley: *Nature*, **318**, 162 (1985).
31) J. Cami, J. Bernard-Salas, E. Peeters and S. E. Malek: *Science,* **329**, 1180 (2010).
32) K. Sellgren, M. W. Werner, J. G. Ingalls, J. D. T. Smith, T. M. Carleton and J. Christine: *ApJL,* **722**, L54 (2010).
33) Y. Zhang, & S. Kwok: *ApJ*, **730**, 126 (2011).
34) O. Bern´e & A. G. G. M.Tielens: *PNAS*, **109**, 4010 (2012) Also, Bern´e, N. L. J. Cox, G. Mulas & C. Joblin: *A&A*, **605**, L1 (2017).
35) D. A. Garc´ıa-Hern´andez, A. Manchado, P. Garc´ıa-Lario, et al.: *ApJL*, **724**, L39 (2010).
36) M. C. Martin, D. Koller, and L. Mihaly: Phys. Rev. **B47**, 14607 (1993).
37) J. Fabian: *Phys. Rev.* **B53**, 13864 (1996).
38) A. Candian, M. G. Rachid, H. MacIssac, V. N. Staroverov, E. Peeters, and J. Cami: preprint from web site of *ResearchGate* by Allessandra Candians, titled "Searching for stable fullerenes in space with computational chemistry" (2009).
39) H.W. Kroto & K. McKay: *Nature*, **331**, 328 (1988).
40) A. Chuvilin, U. Kaiser, E. Bichoutskaia, N. A. Besley, & A.N. Khlobystov: *Nature Chem.*, **2**, 450 (2010).
41) A. K. Geim & K. S. Novoselov: *Nature Materials,* **6**, 183 (2007).
42) D. A. Garcıa-Hernandez, Rao N. Kameswara & D. L. Lambert: *ApJ*, **729**, 126 (2011).
43) D. A. Garcıa-Hernandez, S. Iglesias, J. A. Acosta-Pulido, et al.: *ApJL,* **737,** L30 (2011).
44) D. A. Garcıa-Hernandez, E. Villaver, P. Garc´ıa-Lario, et al.: *ApJ*, **760**, 107 (2012).
45) C. Duboscq, F. Caldvo, M. Rapacioli, E. Dartois, T. Pino, C. Falvo, and A. Simon: *A& A,* **634,** A62 (2020).
46) N. Ota: *arXiv,* 1412.0009 (2014), Additional data for scaling factor on *arXiv,* 1502.01766, for emission spectrum calculation on *arXiv,* 1703.05931, for SP3 defect mechanism on *arXiv,* 1808.01070.
47) H. Galue, and G. Leines: *Phys. Rev. Lett.,* **119**, 171102 (2017).
48) P. Hohenberg and W. Kohn: *Phys. Rev.*, **136**, B864 (1964).
49) W. Kohn and L. Sham: *Phys. Rev.*, **140**, A1133 (1965).
50) A. Becke: *J. Chem. Phys.*, **98**, 5648 (1993).
51) M. Frisch, G. Trucks, H. Schlegel et al: Gaussian 09 package software, Gaussian Inc. Wallington CT USA (2009).
52) R. Ditchfield, W. Hehre and J. Pople: *J. Chem. Phys.,* **54**,724(1971).
53) B. T. Draine and A. Li: *ApJ,* **551**, 807 (2001).
54) A. Li and B. T. Draine: *ApJ,* **554**, 778 (2001).